\documentclass[10pt,twocolumn,oneside,journal]{IEEEtran}
\usepackage{cite}
\usepackage{subfigure}
\usepackage{graphicx}
\usepackage{amsmath}
\usepackage{txfonts}
\usepackage{times}
\usepackage{latexsym}
\usepackage{graphicx}
\usepackage{bm}
\usepackage{amssymb}
\usepackage[center]{caption2}
\usepackage{stfloats}
\usepackage{cases}
\usepackage{array}
\usepackage{setspace}
\usepackage{fancyhdr}
\usepackage{balance} 
\usepackage{flushend} 

\usepackage{algorithm}
\usepackage{algorithmic}
\usepackage{CJK}
\usepackage{xcolor}

\allowdisplaybreaks[4]

\newcaptionstyle{mystyle1}{%
  \centering TABLE  \captiontext \par}
\captionstyle{mystyle1}
\newcaptionstyle{mystyle2}{%
  \captionlabel $.$ \: \captiontext \par}
\captionstyle{mystyle2}
\newcaptionstyle{mystyle3}{%
  \captionlabel $.$ \: \captiontext \par}
\captionstyle{mystyle3}

\renewcommand{\QED}{\QEDopen}

\begin{document}

\title{\huge Decentralized Asynchronous Coded Caching in Fog-RAN}


\author{Wenlong~Huang,
Yanxiang~Jiang,~\IEEEmembership{Member,~IEEE},
Mehdi~Bennis,~\IEEEmembership{Senior~Member,~IEEE},  Fu-Chun~Zheng,~\IEEEmembership{Senior~Member,~IEEE},
Haris Gacanin,~\IEEEmembership{Member,~IEEE},
and Xiaohu You ~\IEEEmembership{Fellow,~IEEE}
\thanks{This work is accepted by IEEE VTC 2018 FALL.}
\thanks{W. Huang and Y. Jiang are with the National Mobile Communications Research Laboratory, Southeast University, Nanjing 210096, China,
the State Key Laboratory of Integrated Services Networks, Xidian University, Xi'an 710071, China,
and the Key Laboratory of Wireless Sensor Network $\&$ Communication, Shanghai Institute of Microsystem and Information Technology,
Chinese Academy of Sciences, Shanghai 200050, China. (e-mail: yxjiang@seu.edu.cn)}
\thanks{M. Bennis is with the Centre for Wireless Communications, University of Oulu, Oulu 90014, Finland. (e-mail: bennis@ee.oulu.fi)}
\thanks{F. Zheng is with the School of Electronic and Information Engineering, Harbin Institute of Technology, Shenzhen 518055, China, and the National Mobile Communications Research Laboratory, Southeast University, Nanjing 210096, China. (e-mail: fzheng@ieee.org)}
\thanks{H. Gacanin is with the Nokia Bell Labs, Antwerp 2018, Belgium. (e-mail:haris.gacanin@nokia-bell-labs.com)}
\thanks{X. You is with the National Mobile Communications Research Laboratory, Southeast University, Nanjing 210096, China. (e-mail:xhyu@seu.edu.cn)}
}

\maketitle

\begin{abstract}
In this paper, we investigate asynchronous coded caching in fog radio access networks (F-RAN).
To minimize the fronthaul load, the encoding set collapsing rule and  encoding set partition method are proposed to establish the relationship between the coded-multicasting contents in asynchronous and synchronous coded caching.
Furthermore, a decentralized asynchronous coded caching scheme is proposed, which provides asynchronous and synchronous transmission methods for different delay requirements.
The simulation results show that our proposed scheme creates considerable coded-multicasting opportunities  in asynchronous request scenarios.
\end{abstract}


\section{Introduction}

With the rapid proliferation of smart devices and mobile application services, wireless networks have been suffering an unprecedented data traffic pressure in recent years, especially at peak-traffic moments.
Fog radio access networks (F-RAN), which can effectively reduce the data traffic pressure by placing popular contents closer to users, have been receiving significant attention from both industry and academia.
In F-RAN, fog access points (F-APs) are distributed at the edges and connected to the cloud server through fronthaul links.
F-APs can use edge computing and caching resources to bring users better quality of experience \cite{Zhang}.
Meanwhile,
as just a few popular content sources account for most of the traffic load,  edge caching has become a trend for content delivery \cite{Bastug, Wang}. Moreover, coded caching was firstly proposed in \cite{Maddah-Ali1} and \cite{Maddah-Ali2} by encoding the delivered contents to further reduce network congestion.

The main idea of coded caching is that the contents stored in the caches  can be used to create coded-multicasting opportunities, such that a single coded-multicasting content transmitted by the cloud server can be useful to a large number of users simultaneously even though  they are not requesting the same content. In \cite{Maddah-Ali1}, Maddah-Ali and Niesen  proposed a centralized coded caching scheme, in which the centrally coordinated placement phase needs the knowledge of the number of active users in the delivery phase. A decentralized coded caching scheme was further proposed in \cite{Maddah-Ali2}, which achieves order-optimal memory-load tradeoff in the asymptotic regime of infinite file size.
The authors of \cite{Niesen1} presented  a strategy which partitions the file library into subsets of approximately uniform request probability and applies the strategy in \cite{Maddah-Ali2} to each subset. The authors of \cite{Hachem} studied the case that the file popularity has multiple different levels.
A scheme consisting of a random popularity-based caching policy and chromatic-number index coding delivery was proposed in \cite{Ji1}, which was proven to be order optimal in terms of average rate.
 All the  schemes in \cite{Maddah-Ali1, Maddah-Ali2, Niesen1, Hachem, Ji1} considered the coded caching problem for the case that user requests are  synchronous, i.e., synchronous coded caching. However,
 user demands for contents is typically asynchronous\cite{Maddah-Ali3} in reality.
 The asynchronous request case was first mentioned in \cite{Maddah-Ali2}, and the authors applied the proposed decentralized synchronous coded caching scheme to an asynchronous request scenario in a simple way.
In \cite{Ghasemi},
the authors proposed a linear programming  formulation for the offline case that the server knows the arrival time before starting transmission. As for the online case that user requests are revealed to the server over time, they considered the situation that users do not have deadlines but wish to minimize the overall completion time.

Motivated by the aforementioned discussions, it is important to study the coded caching problem when  user requests are  asynchronous, i.e.,   asynchronous coded caching.
We consider the online case with a given maximum request delay to reduce the worst-case load of the fronthaul links in F-RAN.
We propose a decentralized asynchronous coded caching scheme, which effectively exploits the coded-multicasting opportunities.
Our proposed scheme is applicable for various asynchronous request scenarios by  providing asynchronous and synchronous transmission methods, which are chosen according to different  delay requirements.

\section{System Model}

Consider the F-RAN where there are  $K$ F-APs and each F-AP serves multiple users.
Assume that the users request contents asynchronously during the time interval $\left( {0,T} \right]$.
Let ${\cal K} = \left\{ {1,2,\cdots,k,\cdots,K} \right\}$  denote the index set of the considered $K$ F-APs.
The cloud server has access to a content library of $N$  files, denoted by ${W_1},{W_2},\cdots,{W_N}$.
Let ${\cal N} = \left\{ {1,2,\cdots,n,\cdots,N} \right\}$  denote the index set of $N$ files with $N \ge K$.
Assume that the size of each file  is $F$  bits and the content library has a uniform popularity distribution.
For each F-AP, only one of its served users requests one file 
during the time interval $\left( {0,T} \right]$, while the F-AP informs the cloud server of the request immediately.
For description convenience, we say $K$ F-APs request contents asynchronously during the considered time interval $\left( {0,T} \right]$, where each F-AP only requests one file.
 Each F-AP has an isolated normalized (by $F$) cache size $M$ for $0 < M < N$.

In the placement phase, the F-APs are given access to the content library.
 By using the same setting of this phase in \cite{Maddah-Ali2}, F-AP $k$ is able to store its cache content $Z_k$ using the content library independently of the other F-APs, i.e., in a decentralized manner.  Let $\phi {}_k$ denote the caching function  of F-AP $k $ which maps the content library into the corresponding cache content, i.e.,  ${Z_k} = \phi {}_k\left( {{W_1},{W_2},\cdots,{W_N}} \right)$.
Note that the size of ${Z_k}$ is  $MF$ bits.

In the delivery phase,  the cache contents of all the  F-APs are first informed to  the cloud server  and  then noted as  cache records by the cloud server.
Without loss of generality, assume that
the time interval  $\left( {0,T} \right]$ is divided into $B$ time slots with $B \ge 2$.
Let $\Delta t = {T \mathord{\left/
 {\vphantom {T B}} \right.
 \kern-\nulldelimiterspace} B}$ denote the time duration of each time slot.
Then time
slot $b \in \left\{ {1,2,\cdots,B} \right\}$  represents the time interval $\left( {\left( {b - 1} \right)\Delta t,b\Delta t} \right]$.
Let ${{\cal U}_b} \subseteq {\cal K}$  denote the index set of the F-APs whose requests arrive during time slot  $b$ with ${{\cal U}_b} \ne \varnothing $.
It is  assumed that the cloud server is informed of the requests of the F-APs in ${\cal U}_b$ during time slot $b$  and processes them in a unified manner, i.e., the cloud server transmits the coded-multicasting
content to all the $K$ F-APs through the fronthaul links at the end of each time slot for the online case.
Suppose that the maximum request delay  is $\Delta b \in \left\{ {1,2,\cdots,B} \right\}$ time slots, where $\Delta b$ denotes
the maximum number of
time slots it takes for an F-AP to recover its requested file.
 Note that we do not consider the time that it takes for the cloud server to transmit the corresponding contents and the time that it takes for each F-AP to transmit the recovered file to the served user. Therefore, the cloud server can fulfill the requests of the F-APs in ${\cal U}_b$ by the end of the time slot $b + \Delta b - 1$.

Let  ${d_k} \in {\cal N}$ denote the index of the file requested by  F-AP $k$ during $\left( {0,T} \right]$, and ${{\boldsymbol{d}}_b} \in {{\cal N}^{\left| {{{\cal U}_b}} \right|}}$ denote the request vector of the corresponding  F-APs in ${\cal U}_b$.
Let $\psi_b$ denote the encoding function of the cloud server at the end of time slot $b$, which maps the files ${W_1},{W_2},\cdots,{W_N}$, the cache contents ${Z_1},{Z_2},\cdots,{Z_K}$, and the requests ${{\boldsymbol{d}}_b}$
 to the coded-multicasting content ${X_b} \buildrel \Delta \over = \psi_b \left( {{W_1}, {W_2}, \cdots,{W_N},{Z_1},{Z_2},\cdots,{Z_K},{{\boldsymbol{d}}_b}} \right)$.
 Let $\theta _{k} $ denote the decoding function of F-AP $k$, which maps the received  coded-multicasting contents ${X_1},{X_2},\cdots,{X_B}$, the cache content $Z_k$, and the request ${{d_k}}$ to the estimate ${\hat W_{d_k}} = {\theta _{k}}\left( {{X_1},{X_2}, \cdots ,{X_B},{Z_k},{d_k}} \right)$ of the requested file ${{W_{{d_k}}}}$ of F-AP $k$. Each F-AP should be able to recover its requested file successfully from its cached content and the received coded-multicasting content, and then transmit  it to the served user. An asynchronous coded caching scheme is said to be feasible if and only if the following condition is satisfied:
\[\mathop {\lim }\limits_{F \to \infty } \mathop {\max }\limits_{{{\boldsymbol{d}}_1},{{\boldsymbol{d}}_2}, \cdots ,{{\boldsymbol{d}}_B}} \mathop {\max }\limits_{k \in {\cal K}} P\left( {{{\hat W}_{d_k}} \ne {W_{{d_k}}}} \right) = 0.\]
Note that the worst-case propability of error over all possible requests ${{\boldsymbol{d}}_1},{{\boldsymbol{d}}_2},\cdots,{{\boldsymbol{d}}_B}$ with infinite $F$ is maximized in the above condition.
The  objective  of this paper is to  find a feasible asynchronous coded caching scheme to minimize the worst-case   normalized  fronthaul load (over all possible requests ${{\boldsymbol{d}}_1},{{\boldsymbol{d}}_2},\cdots,{{\boldsymbol{d}}_B}$) in the delivery phase with a given $\Delta b$.

\section{The Proposed
Decentralized Asynchronous Coded Caching Scheme}

In this section, we first propose the encoding set collapsing rule.
 Then, we present the encoding set partition method. Finally,
  a novel decentralized asynchronous coded caching scheme is proposed to minimize the fronthaul load.

\subsection{The Proposed Encoding Set Collapsing Rule}

Asynchronous coded caching and synchronous coded caching are supposed to be  under the same condition when their system parameters $M$, $K$, and $N$ are the same.
As in the conventional synchronous coded caching schemes under the same condition, such as the  Maddah-Ali-Niesen's decentralized scheme\cite{Maddah-Ali2},  ${\cal S} \subseteq {\cal K}$ for any $s = \left| {\cal S} \right| \in \left\{ {1,2,\cdots,K} \right\}$ is called an encoding set if
a single coded-multicasting content can be useful to the F-APs in $\cal S$ simultaneously.
It can be  observed that the subset of $\cal S$ is also an encoding set.
In order to differentiate the same subfile in asynchronous  and synchronous coded caching,
let $W_{k,{\cal S}}^{\rm a}$ and $W_{k,{\cal S}}^{\rm s}$ denote the bits of the file requested by F-AP $k$ cached exclusively at the F-APs in ${\cal S}$ for asynchronous and synchronous coded caching, respectively.

Consider that the requests of the F-APs in ${{\cal K}}\backslash {{\cal U}_{{1}}}$  have not arrived yet during time slot $1$. Assume that  ${{\cal U}_1} \cap {\cal S} \ne \varnothing $ and  the requests
in ${{\boldsymbol{d}}_1}$
should be fulfilled at the end of time slot 1.
The cloud server needs to transmit a coded-multicasting content which is useful to the F-APs in ${{\cal U}_1} \cap {\cal S}$ at the end of time slot 1.
Thus, we say that the encoding set $\cal S$ in synchronous coded caching collapses into a subset of $\cal S$, i.e., ${{\cal U}_1} \cap {\cal S}$, for transmitting the corresponding coded-multicasting content in asynchronous coded caching.
Recall that by applying the scheme in \cite{Maddah-Ali2}, the coded-multicasting content that the cloud server
transmits for $\cal S$ in synchronous coded caching is
${ \oplus _{k \in {\cal S}}}W_{k,{\cal S}\backslash \left\{ k \right\}}^{\rm s}$,
 where $ \oplus $  denotes bitwise XOR operation.
Accordingly, the cloud server transmits
${ \oplus _{k \in \left( {{\cal S} \cap {{\cal U}_{{1}}}} \right)}}{W_{k,\left( {{\cal S} \cap {{\cal U}_{{1}}}} \right)\backslash \left\{ k \right\}}^{\rm a}}$ at the end of time slot 1.
According to the above discussions, it is evident that there exists some relationship, called encoding set collapsing rule, between the coded-multicasting contents in asynchronous and synchronous coded caching.

During time slot $b$, let ${\cal U}_y$ and ${\cal U}_n$ denote the index sets of the F-APs  from which the requests have arrived  and  not arrived, respectively.
For any ${{\cal S}_1} \subseteq {{\cal U}_y}$  and ${{\cal S}_2} \subseteq {{\cal U}_n}$, the encoding set ${\cal S} = {{\cal S}_1} \cup {{\cal S}_2}$ in synchronous coded caching collapses into ${\cal S}_1$ in asynchronous coded caching.
Accordingly,
  ${ \oplus _{k \in \left( {{{\cal S}_1} \cup {{\cal S}_2}} \right)}}W_{k,\left( {{{\cal S}_1} \cup {{\cal S}_2}} \right)\backslash \left\{ k \right\}}^{\rm s} $ collapses into ${ \oplus _{k \in {{\cal S}_1}}}{W_{k,\left( {{{\cal S}_1} \cup {{\cal S}_2}} \right)\backslash \left\{ k \right\}}^{\rm a}}$, which will be practically transmitted by the cloud server  at the end of time slot $b$ in asynchronous coded caching.

\subsection{The Proposed Encoding Set Partition Method}

Utilizing our proposed encoding set collapsing rule, we now consider what contents are transmitted in asynchronous coded caching.
In order to fulfill the requests of the F-APs
in asynchronous coded caching, $\cal S$ may need to be collapsed into a subset of $\cal S$ many times to transmit the corresponding  content at the end of different time slots.
For a given $\Delta b$,  only the requests of the F-APs in $\mathop  \cup \nolimits_{i = \max \left\{ {1,b - \Delta b + 1} \right\}}^b {{\cal U}_i}$, referred to as active F-APs, need to be fulfilled during time slot $b$.
Let ${{\cal U}^ {\rm a}} = \mathop  \cup \nolimits_{i = \max \left\{ {1,b - \Delta b + 1} \right\}}^b {{\cal U}_i}$ denote the index set of the active F-APs during time slot $b$.
Then, only the files requested by the F-APs in ${{\cal U}^ {\rm a}}$ can be encoded with each other, which means that $\cal S$ collapses into ${\cal S} \cap {{{\cal U}^ {\rm a}}}$.
Moreover, to minimize the fronthaul load, it is equivalent to partition $\cal S$ into
the minimum number of nonoverlapping subsets for transmission in asynchronous coded caching.

\begin{algorithm}[!t]
\footnotesize
\caption{The Proposed Encoding Set Partition Method}
\label{partitionMethod}
\begin{algorithmic}[1]

\STATE Initialize $i $, $\beta$, $\gamma$.
\WHILE{${\cal S} \ne \varnothing $}
\STATE $i =  i + 1$.
\WHILE{${{\cal U}_{\beta + 1}} \cap {\cal S} = \varnothing $}
\STATE $\beta =  \beta + 1$.
\ENDWHILE

\IF{$\gamma - \beta \ge  \Delta b$}
\STATE ${\cal S}_i  = {\cal S} \cap \left( {\mathop  \cup \nolimits_{b = \beta + 1}^{\beta + \Delta b} {{\cal U}_b}} \right)$,
\STATE $\beta = \beta + \Delta b$.
\ELSE
\STATE ${\cal S}_i  = {\cal S} \cap \left( {\mathop  \cup \nolimits_{b = \beta + 1}^\gamma {{\cal U}_b}} \right)$.
\ENDIF
\STATE ${\cal S} = {\cal S}\backslash {\cal S}_i$.
\ENDWHILE
\end{algorithmic}
\end{algorithm}

Let $\left( {\beta \Delta t,\gamma \Delta t} \right]$ denote the active time interval  of ${\cal S}$ if $\left( {\left( {\mathop  \cup \nolimits_{b = 1}^\beta {{\cal U}_b}} \right) \cup \left( {\mathop  \cup \nolimits_{b = \gamma + 1}^B {{\cal U}_b}} \right)} \right) \cap {\cal S} = \varnothing $ and ${{\cal U}_b} \cap {\cal S} \ne \varnothing$ for
$b = \beta  + 1$ and $  \gamma $, where
 $\beta$ and $\gamma$ are integers with $0 \le \beta  < \gamma \le {B}$.
Let
${\cal S}_i$ denote the $i$th encoding subset that ${\cal S}$ is partitioned into for a given $\Delta b$.
The detailed encoding set partition method is presented in Algorithm \ref{partitionMethod}.

\subsection{The Proposed Asynchronous Coded Caching Scheme}

 According to the above discussions,
  we propose the decentralized asynchronous coded caching scheme which implements the  encoding set partition method.
 In the placement phase, each F-AP randomly selects  ${M}F/{N}$ bits of each file with uniform probability and fetch them to fill its cache, which is the same as  the Maddah-Ali-Niesen's decentralized synchronous coded caching scheme.
Note that the placement procedure does not require any coordination and can be operated in a decentralized manner. More specifically, our proposed scheme can operate in the placement phase with an unknown number of F-APs.

In the delivery phase, we propose asynchronous and synchronous transmission methods for the online case, which can be chosen by the cloud server.
Note that asynchronous or synchronous here means that the cloud server transmits the coded-multicasting contents asynchronously or synchronously.

\subsubsection{Asynchronous Transmission Method}

 When $\Delta b < B$,
 the asynchronous transmission method is chosen.
 As the requests  in ${{\boldsymbol{d}}_b}$
  need to be fulfilled by the end of the time slot $b + \Delta b - 1$,
  we should try to fulfill the requests at the end of   time  slot $b + \Delta b - 1$ so that sufficient  coded-multicasting opportunities can be created.
If $\Delta b > 1$, no contents are transmitted
at the end of  time slot $1, 2,\cdots ,\Delta b - 1$. Then,  only the  requests of the F-APs in ${{\cal U}^ {\rm a}}$
 need to be fulfilled by the cloud server at the end of the time slots between $\Delta b - 1 $ and $ B$.
 For description convenience, we say the subfile ${W_{k,{\cal S}}}$ is of type $s$  with  $s \in \left\{ {0,1,\cdots,K} \right\}$.
  Thus, the cloud server transmits a single coded-multicasting content for $\cal S$ by encoding the subfiles of type $s-1$\cite{Maddah-Ali2}.
  At the end of this time slot, the cloud server firstly partitions each file $W_n$ into $K + 1$ types of nonoverlapping subfiles. Note that there are
  \begin{small}
$\left( {\begin{array}{*{20}{c}}
K\\
s
\end{array}} \right)$
\end{small}
subfiles of the form $\left( {{W_{n,{\cal S}}^{\rm a}}:{\cal S} \subseteq {\cal K},\left| {\cal S} \right| = s} \right)$ for each type $s \in \left\{ {0,1,\cdots,K} \right\}$,  whose sizes are calculated based on the updated cache records.
For any $s$, define $\underline \chi   = \max \left\{ {1,s + \left| {{{\cal U}_{b - \Delta b + 1}}} \right| - \left| {\cal K} \right|} \right\}$ and $\overline \chi   = \min \left\{ {s,\left| {{{\cal U}_{b - \Delta b + 1}}} \right|} \right\}$. Consider any $\chi  \in \left\{ {\underline \chi  ,\underline \chi   + 1, \cdots ,\overline \chi  } \right\}$.
Focus on an encoding subset   ${{\cal S}^1} \subseteq {{\cal U}_{b - \Delta b + 1}}$  with $\left| {{{\cal S}^1}} \right| = \chi $ and an encoding subset ${{\cal S}^2} \subseteq {\cal K} \backslash {{\cal U}_{b - \Delta b + 1}}$ with $\left| {{{\cal S}^2}} \right| = s - \chi $.
Recall that the  F-APs in $\left( {{{\cal S}^1} \cup {{\cal S}^2}} \right)\backslash \left\{ k \right\}$ share a subfile which is not available at the cache content ${{Z_k}}$ and requested by F-AP  $k \in \left( {{{\cal S}^1} \cup {{\cal S}^2}} \right) $.
For any ${\cal S}^1$ and  ${\cal S}^2$ with any $s$, in order to avoid some subfiles are transmitted repeatedly,
 no contents are transmitted
if ${W_{k,\left( {{{\cal S}^1} \cup {{\cal S}^2}} \right)\backslash \left\{ k \right\}}^{\rm a}} = \varnothing $ for $k \in \left( {\left( {{{\cal S}^1} \cup {{\cal S}^2}} \right) \cap {{{\cal U}^ {\rm a}}}} \right)$.
Otherwise, the cloud server
  transmits the coded-multicasting content as follows:
 \[{ \oplus _{k \in \left( {\left( {{{\cal S}^1} \cup {{\cal S}^2}} \right) \cap {{{\cal U}^ {\rm a}}}} \right)}}{W_{k,\left( {{{\cal S}^1} \cup {{\cal S}^2}} \right)\backslash \left\{ k \right\}}^{\rm a}}.\]

 After the transmission is completed,
  each F-AP in ${{\cal U}_{b - \Delta b + 1}}$ recovers  the desirable subfiles of its requested file.
 Then, each F-AP in ${{\cal U}_{b - \Delta b + 1}}$ transmits  the recovered subfiles and the  corresponding subfiles available in its cache to its served user immediately; thus the user can recover the desirable file.
 While each F-AP in ${{{\cal U}^ {\rm a}}}\backslash {{\cal U}_{b - \Delta b + 1}}$ also  recovers the corresponding desirable subfiles, and then  transmit them to its served user at this time.
In addition, the cloud server needs to update the cache records of the active F-APs
  by adding a record of the subfiles recovered by each F-AP in ${{\cal U}^ {\rm a}}$  at the end of this time slot as its cache content.
  Note that updating the cache records has no influence on the cache contents of the F-APs, which stay unchanged in the delivery phase. The cache records is used to help the cloud server identify  whether the subfile to be transmitted is $\varnothing$ in real time before transmission.

\begin{algorithm}[!t]
\footnotesize
\setcounter{algorithm}{1}
\caption{The Proposed Asynchronous Coded Caching Scheme}
\label{alg}
\begin{algorithmic}[1]
\STATE PLACEMENT
\FOR{ $k \in {\cal K}, n \in {\cal N}$}
\STATE F-AP $k$  independently caches a subset of ${MF}/{N}$ bits of file $W_n$, chosen uniformly at random.
\ENDFOR
\\-------------------------------------------------------------
\STATE DELIVERY
\STATE Initialize ${{\cal U}^ {\rm a}} = \varnothing$, $b  = 1$.
\WHILE{$b \le B $}

\IF{$ \Delta b < B$}

\IF{$b \le \Delta b - 1$}
\STATE ${{\cal U}^ {\rm a}} = {{\cal U}^ {\rm a}} \cup {{\cal U}_b}$,
\STATE At the end of time slot $b$,  no contents are transmitted.

\ELSIF{$\Delta b - 1 < b < {B}$}
\STATE ${\cal U}^ {\rm a} = {\cal U}^ {\rm a} \cup {\cal U}_b$.
\FOR{$s = \left| {\cal K} \right|,\left| {\cal K} \right| - 1,\cdots,1$}
\FOR{$\chi  = \max \left\{ {1,s + \left| {{{\cal U}_{b - \Delta b + 1}}} \right| - \left| {\cal K} \right|} \right\}:\min \left\{ {s,\left| {{{\cal U}_{b - \Delta b + 1}}} \right|} \right\}$}
    \FORALL{${\cal S}^1 \subseteq {{\cal U}_{b - \Delta b + 1}},{\cal S}^2 \subseteq {{\cal K}} \backslash {{\cal U}_{b - \Delta b + 1} }:\left| {{\cal S}^1} \right| = \chi ,\left| {{\cal S}^2} \right| = s - \chi$}
        \STATE At the end of time slot $b$,  no contents are transmitted if ${W_{k,\left( {{{\cal S}^1} \cup {{\cal S}^2}} \right)\backslash \left\{ k \right\}}^{\rm a}} = \varnothing $ for  $k \in \left( {\left( {{{\cal S}^1} \cup {{\cal S}^2}} \right) \cap {{\cal U}^ {\rm a}}} \right)$;
        Otherwise, the cloud server sends
        ${ \oplus _{k \in \left( {\left( {{{\cal S}^1} \cup {{\cal S}^2}} \right) \cap {{\cal U}^ {\rm a}}} \right)}}{W_{k,\left( {{{\cal S}^1} \cup {{\cal S}^2}} \right)\backslash \left\{ k \right\}}^{\rm a}}$.
    \ENDFOR
\ENDFOR
\ENDFOR
    \STATE ${{\cal U}^ {\rm a}} = {{\cal U}^ {\rm a}} \backslash {{\cal U}_{b - \Delta b+ 1}}$;

\ELSE
\STATE ${\cal U}^ {\rm a} = {\cal U}^ {\rm a} \cup {\cal U}_b$.
\FOR{$s = \left| {\cal K} \right|,\left| {\cal K} \right| - 1,\cdots,1$}
\FOR{$\chi  = \max \left\{ {1,s + \left| {{{\cal U}^ {\rm a}}} \right| - \left| {\cal K} \right|} \right\}:\min \left\{ {s,\left| {{{\cal U}^ {\rm a}}} \right|} \right\}$}
    \FORALL{${\cal S}^1 \subseteq {{\cal U}^ {\rm a}},{\cal S}^2 \subseteq {{\cal K}} \backslash {{\cal U}^ {\rm a} }:\left| {{\cal S}^1} \right| = \chi ,\left| {{\cal S}^2} \right| = s - \chi$}
        \STATE At the end of time slot $B$,
        the cloud server sends
        ${ \oplus _{k \in {\cal S}^1}}{W_{k,\left( {{{\cal S}^1} \cup {{\cal S}^2}} \right)\backslash \left\{ k \right\}}^{\rm a}}.$

    \ENDFOR
\ENDFOR
\ENDFOR
\ENDIF

\ELSE
\IF{$b \le B - 1$}
\STATE At the end of time slot $b$,  no contents are transmitted.
\ELSE
\FOR{$s = \left| {\cal K} \right|,\left| {\cal K} \right| - 1,\cdots,1$}
    \FORALL{${\cal S} \subseteq {{\cal K}}:\left| {\cal S} \right| = s$}
        \STATE At the end of time slot $B$, the cloud server sends ${ \oplus _{k \in {\cal S}}}{W_{k,{\cal S}\backslash \left\{ k \right\}}^{\rm a}}$.
    \ENDFOR
\ENDFOR
\ENDIF

\ENDIF

\STATE $b = b + 1$.
\ENDWHILE

\end{algorithmic}
\end{algorithm}

 At the end of time slot $B$, all the requests of the F-APs in ${\cal U}^ {\rm a}$ can be fulfilled together. Similarly, define $\underline \chi^\prime   = \max \left\{ {1,s + \left| {{{\cal U}^ {\rm a}}} \right| - \left| {\cal K} \right|} \right\}$ and $\overline \chi^\prime   = \min \left\{ {s,\left| {{{\cal U}^ {\rm a}}} \right|} \right\}$.
 For any ${\cal S}^1 \subseteq {{\cal U}^ {\rm a}}$ of cardinality $\left| {{{\cal S}^1}} \right| = \chi$ and  ${\cal S}^2 \subseteq {{\cal K}} \backslash {{\cal U}^ {\rm a} }$ of cardinality $\left| {{{\cal S}^2}} \right| = s - \chi $ with any $s$ and $\chi \in \left\{ {\underline \chi^\prime  ,\underline \chi^\prime   + 1, \cdots ,\overline \chi^\prime  } \right\}$,
  the cloud server transmits the coded-multicasting content as follows:
 \[{ \oplus _{k \in {\cal S}^1}}{W_{k,\left( {{{\cal S}^1} \cup {{\cal S}^2}} \right)\backslash \left\{ k \right\}}^{\rm a}},\]
 where all the subfiles ${W_{k,\left( {{{\cal S}^1} \cup {{\cal S}^2}} \right)\backslash \left\{ k \right\}}}$ are assumed to be zero-padded to the number of bits of the longest subfile in the bit-wise XOR operation.
After that, each F-AP in ${{\cal U}^ {\rm a}}$ recovers  the subfiles of its requested file, and then transmits the recovered subfiles and the subfiles available in its cache to its served user; thus the user can recover the desirable file.

\subsubsection{Synchronous Transmission Method}

When $\Delta b = B$, the synchronous transmission method is chosen.
Firstly, no contents are transmitted
at the end of  time slot $1, 2,\cdots ,B - 1$.
At the end of time slot $B$, for all $\cal S$ with any $s$, the cloud server transmits the coded-multicasting content as follows:
  \[{ \oplus _{k \in {\cal S}}}{W_{k,{\cal S}\backslash \left\{ k \right\}}^{\rm a}}.\]
Then, each F-AP transmits all  the subfiles of its requested file to its served user; thus the user can recover the desirable file.

The detailed description of our proposed decentralized asynchronous coded caching scheme is presented in Algorithm \ref{alg}.
Note that $\chi$ is used to ensure that ${{\cal S}^1} \cap {\cal S} \ne \varnothing $, so that the coded-multicasting content transmitted by the cloud server for  $\cal S$ is useful to at least one F-AP in  ${{\cal U}_{b - \Delta b + 1}}$ or ${\cal U}^{\rm a}$.
Also note that the problem setting allows
for vanishing probability of error as $F \to \infty$.


\emph{Example 1:} Assume that $N = 4$, $K = 4$, $M = 2$, $B = 4$,
$T = 4 \ \rm s$, $\Delta t = 1 \ \rm s$, $\Delta b = 2$, ${{\cal U}_b} = \left\{ b \right\}$ and $d_k = k$  in asynchronous coded caching.
It is easy to see that this is the worst-case request.
According to Algorithm \ref{alg}, the coded-multicasting contents transmitted by the cloud server at the end of time slot 2, 3, and 4  are illustrated in Table \ref{res2}, Table \ref{res3}, and Table \ref{res4}, respectively.
Note that  $\varnothing $ indicates no contents are transmitted in the tables.  In addition, subfile ${W_{3,\left\{ {2,4} \right\}}^{\rm a}}$ is actually $\varnothing$ according to the updated cache records.

Consider the same setting as Example 1. Now, we explain how  Algorithm \ref{alg} implements our proposed encoding set partition method.
 Focus on
  ${\cal S} = \left\{ {1,2,3,4} \right\}$.
 Firstly, no contents are transmitted at the end of time slot $1$. At the end of time slot $2$,
  ${W_{1,\left\{ {2,3,4} \right\}}^{\rm a}} \oplus {W_{2,\left\{ {1,3,4} \right\}}^{\rm a}}$ is transmitted with ${{\cal U}^ {\rm a}} = \left\{ {1,2} \right\}$.
 At the end of time slot $3$, the cloud server decides not to transmit ${W_{2,\left\{ {1,3,4} \right\}}^{\rm a}} \oplus {W_{3,\left\{ {1,2,4} \right\}}^{\rm a}}$ with ${{\cal U}^ {\rm a}} = \left\{ {2,3} \right\}$, since ${W_{2,\left\{ {1,3,4} \right\}}^{\rm a}}$ is $\varnothing $
according to the updated cache records.
Thus, no contents are transmitted.
 Finally,
 ${W_{3,\left\{ {1,2,4} \right\}}^{\rm a}} \oplus {W_{4,\left\{ {1,2,3} \right\}}^{\rm a}}$ is transmitted with ${{\cal U}^ {\rm a}} = \left\{ {3,4} \right\}$
 at the end of time slot $4$. Observe that  $W_{1,\left\{ {2,3,4} \right\}}^{\rm s} \oplus W_{2,\left\{ {1,3,4} \right\}}^{\rm s} \oplus W_{3,\left\{ {1,2,4} \right\}}^{\rm s} \oplus W_{4,\left\{ {1,2,3} \right\}}^{\rm s}$ is partitioned into two parts of equal size, i.e., ${W_{1,\left\{ {2,3,4} \right\}}^{\rm a}} \oplus {W_{2,\left\{ {1,3,4} \right\}}^{\rm a}}$ and ${W_{3,\left\{ {1,2,4} \right\}}^{\rm a}} \oplus {W_{4,\left\{ {1,2,3} \right\}}^{\rm a}}$, for transmission in our proposed asynchronous coded caching scheme.

\emph{Remark 1:}
The main innovation of our proposed scheme is
to partition the coded-multicasting contents in synchronous coded caching by using our proposed encoding set partition method, and then transmit the partitioned contents  at the end of different time slots.
Our proposed scheme  can create
as many coded-multicasting opportunities as possible
while the maximum request delay of each F-AP is no more than $\Delta b$ time slots.

\emph{Remark 2:} The asynchronous coded caching problem has been considered in \cite{Ghasemi}. The authors  described their proposed approach based on a system using the centralized synchronous coded caching scheme in \cite{Maddah-Ali1}, while they declared their approach can be applied to general placement schemes.
In this paper, we  propose a decentralized asynchronous coded caching scheme based on a different system model for the online case, which is more applicable for practical scenarios.
Moreover, our proposed scheme can work well for both the online case and offline case.

\begin{table}[!t]
\centering
\setlength{\abovecaptionskip}{0pt}%
\setlength{\belowcaptionskip}{10pt}%
\caption{The contents transmitted during time slot $2$}
\label{res2}
\scalebox{0.9}
{
\begin{tabular}{|c|c|c|c|c|c|}
\hline
$s$ & $\chi$ & ${\cal S}^1$ & ${\cal S}^2$ & ${\cal U}^ {\rm a}$ & Coded-multicasting Content   \\
\hline
4 & 1 & $\left\{ 1 \right\}$ & $\left\{ {2,3,4} \right\}$ & $\left\{ {1,2} \right\}$ & ${W_{1,\left\{ {2,3,4} \right\}}^{\rm a}} \oplus {W_{2,\left\{ {1,3,4} \right\}}^{\rm a}}$  \\
\hline
3 & 1 & $\left\{ 1 \right\}$ & $\left\{ {2,3} \right\}$ & $\left\{ {1,2} \right\}$ & ${W_{1,\left\{ {2,3} \right\}}^{\rm a}} \oplus {W_{2,\left\{ {1,3} \right\}}^{\rm a}}$  \\
\hline
3 & 1 & $\left\{ 1 \right\}$ & $\left\{ {2,4} \right\}$ & $\left\{ {1,2} \right\}$ & ${W_{1,\left\{ {2,4} \right\}}^{\rm a}} \oplus {W_{2,\left\{ {1,4} \right\}}^{\rm a}}$  \\
\hline
3 & 1 & $\left\{ 1 \right\}$ & $\left\{ {3,4} \right\}$ & $\left\{ {1,2} \right\}$ & ${W_{1,\left\{ {3,4} \right\}}^{\rm a}}$  \\
\hline
 2& 1 & $\left\{ 1 \right\}$ & $\left\{ {2} \right\}$ & $\left\{ {1,2} \right\}$ & ${W_{1,\left\{ 2 \right\}}^{\rm a}} \oplus {W_{2,\left\{ 1 \right\}}^{\rm a}}$  \\
\hline
2& 1 & $\left\{ 1 \right\}$ & $\left\{ {3} \right\}$ & $\left\{ {1,2} \right\}$ & ${W_{1,\left\{ 3 \right\}}^{\rm a}} $  \\
\hline
2& 1 & $\left\{ 1 \right\}$ & $\left\{ {4} \right\}$ & $\left\{ {1,2} \right\}$ & ${W_{1,\left\{ 4 \right\}}^{\rm a}}$  \\
\hline
1& 1 & $\left\{ 1 \right\}$ & $\varnothing$ & $\left\{ {1,2} \right\}$ & ${W_{1,\varnothing }^{\rm a}}$  \\
\hline
\end{tabular}}
\end{table}

\begin{table}[!t]
\centering
\setlength{\abovecaptionskip}{0pt}%
\setlength{\belowcaptionskip}{10pt}%
\caption{The contents transmitted during time slot $3$}
\label{res3}
\scalebox{0.9}
{
\begin{tabular}{|c|c|c|c|c|c|}
\hline
$s$ & $\chi$ & ${\cal S}^1$ & ${\cal S}^2$ & ${\cal U}^ {\rm a}$ & Coded-multicasting Content   \\
\hline
4 & 1 & $\left\{ 2 \right\}$ & $\left\{ {1,3,4} \right\}$ & $\left\{ {2,3} \right\}$ & $\varnothing $  \\
\hline
3 & 1 & $\left\{ 2 \right\}$ & $\left\{ {1,3} \right\}$ & $\left\{ {2,3} \right\}$ & $\varnothing $ \\
\hline
3 & 1 & $\left\{ 2 \right\}$ & $\left\{ {1,4} \right\}$ & $\left\{ {2,3} \right\}$ & $\varnothing $  \\
\hline
3 & 1 & $\left\{ 2 \right\}$ & $\left\{ {3,4} \right\}$ & $\left\{ {2,3} \right\}$ & ${W_{2,\left\{ {3,4} \right\}}^{\rm a}} \oplus {W_{3,\left\{ {2,4} \right\}}^{\rm a}}$  \\
\hline
 2& 1 & $\left\{ 2 \right\}$ & $\left\{ {1} \right\}$ & $\left\{ {2,3} \right\}$ & $\varnothing $  \\
\hline
2& 1 & $\left\{ 2 \right\}$ & $\left\{ {3} \right\}$ & $\left\{ {2,3} \right\}$ & ${W_{2,\left\{ 3 \right\}}^{\rm a}} \oplus {W_{3,\left\{ 2 \right\}}^{\rm a}}$  \\
\hline
2& 1 & $\left\{ 2 \right\}$ & $\left\{ {4} \right\}$ & $\left\{ {2,3} \right\}$ & ${W_{2,\left\{ 4 \right\}}^{\rm a}}$  \\
\hline
1& 1 & $\left\{ 2 \right\}$ & $\varnothing$ & $\left\{ {2,3} \right\}$ & ${W_{2,\varnothing }^{\rm a}}$  \\
\hline
\end{tabular}}
\end{table}

\begin{table}[!t]
\centering
\setlength{\abovecaptionskip}{0pt}%
\setlength{\belowcaptionskip}{10pt}%
\caption{The contents transmitted during time slot $4$}
\label{res4}
\scalebox{0.9}
{
\begin{tabular}{|c|c|c|c|c|c|}
\hline
$s$ & $\chi$ & ${\cal S}^1$ & ${\cal S}^2$ & ${\cal U}^ {\rm a}$ & Coded-multicasting Content  \\
\hline
4 & 2 & $\left\{ {3,4} \right\}$ & $\left\{ {1,2} \right\}$ & $\left\{ {3,4} \right\}$ & ${W_{3,\left\{ {1,2,4} \right\}}^{\rm a}} \oplus {W_{4,\left\{ {1,2,3} \right\}}^{\rm a}}$  \\
\hline
3 & 1 & $\left\{ 3 \right\}$ & $\left\{ {1,2} \right\}$ & $\left\{ {3,4} \right\}$ & ${W_{3,\left\{ {1,2} \right\}}^{\rm a}}$ \\
\hline
3 & 1 & $\left\{ 4 \right\}$ & $\left\{ {1,2} \right\}$ & $\left\{ {3,4} \right\}$ & ${W_{4,\left\{ {1,2} \right\}}^{\rm a}}$  \\
\hline
3 & 2 & $\left\{ {3,4} \right\}$ & $\left\{ {1} \right\}$ & $\left\{ {3,4} \right\}$ & ${W_{3,\left\{ {1,4} \right\}}^{\rm a}} \oplus {W_{4,\left\{ {1,3} \right\}}^{\rm a}}$  \\
\hline
3 & 2 & $\left\{ {3,4} \right\}$ & $\left\{ {2} \right\}$ & $\left\{ {3,4} \right\}$ & ${W_{3,\left\{ {2,4} \right\}}^{\rm a}}\left( \varnothing  \right) \oplus {W_{4,\left\{ {2,3} \right\}}^{\rm a}}$  \\
\hline
 2& 1 & $\left\{ 3 \right\}$ & $\left\{ {1} \right\}$ & $\left\{ {3,4} \right\}$ & ${W_{3,\left\{ 1 \right\}}^{\rm a}}$  \\
\hline
2& 1 & $\left\{ 3 \right\}$ & $\left\{ {2} \right\}$ & $\left\{ {3,4} \right\}$ & $\varnothing$  \\
\hline
2& 1 & $\left\{ 4\right\}$ & $\left\{ {1} \right\}$ & $\left\{ {3,4} \right\}$ & ${W_{4,\left\{ 1 \right\}}^{\rm a}}$  \\
\hline
2& 1 & $\left\{ 4 \right\}$ & $\left\{ {2} \right\}$ & $\left\{ {3,4} \right\}$ & ${W_{4,\left\{ 2 \right\}}^{\rm a}}$  \\
\hline
2& 2 & $\left\{ {3,4} \right\}$ & $\varnothing$ & $\left\{ {3,4} \right\}$ & ${W_{3,\left\{ 4 \right\}}^{\rm a}} \oplus {W_{4,\left\{ 3 \right\}}^{\rm a}}$  \\
\hline
1& 1 & $\left\{ 3 \right\}$ & $\varnothing$ & $\left\{ {3,4} \right\}$ & ${W_{3,\varnothing }^{\rm a}}$  \\
\hline
1& 1 & $\left\{ 4 \right\}$ & $\varnothing$ & $\left\{ {3,4} \right\}$ & ${W_{4,\varnothing }^{\rm a}}$  \\
\hline
\end{tabular}}
\end{table}

\section{Simulation Results}

    In this section, the performance of our  proposed decentralized asynchronous coded caching scheme is evaluated via simulations.
We adopt  the Maddah-Ali-Niesen's decentralized synchronous coded caching scheme and the uncoded caching scheme as  the baseline schemes.
In our simulations, the number of the F-APs whose requests arrive  during time slot $b$, i.e., $\left| {{{\cal U}_b}} \right|$, is  random.
Other parameters are set as follows: $F = 1 \  \rm Gb$, $N = 100$, $K = 10$, $T = 10 \ \rm s$, $B = 5$.

\begin{figure}[!t]
\centering 
\includegraphics[width=0.45\textwidth]{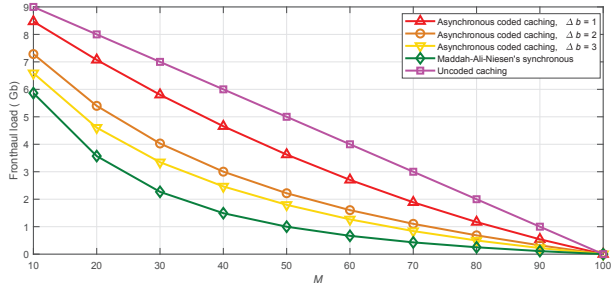}
\caption{ Fronthaul load versus $M$.} \label{cache_load}
\end{figure}

In Fig. \ref{cache_load}, we show the effect of the normalized cache size of each F-AP, i.e., $M$, on the fronthaul load of each scheme for different $\Delta b$.  As shown, our  proposed  scheme can create considerable coded-multicasting opportunities compared with the uncoded caching scheme.
Moreover, the fronthaul load decreases and its slope increases when $M$ increases, which is the same as  the Maddah-Ali-Niesen's decentralized synchronous coded caching scheme.

\begin{figure}[!t]
\centering 
\includegraphics[width=0.45\textwidth]{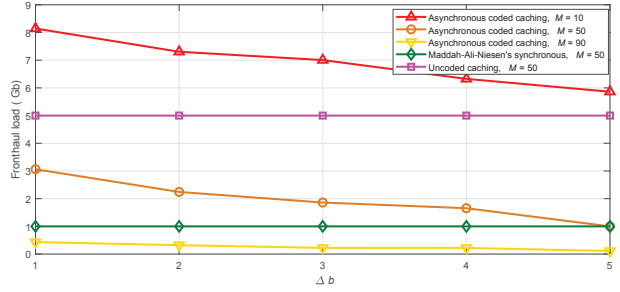}
\caption{ Fronthaul load versus $\Delta b$ for varying cache sizes. } \label{joint}
\end{figure}

In Fig. \ref{joint}, we show how $\Delta b$  affects the fronthaul load of  each scheme for varying cache sizes.
As shown,  the fronthaul load of our proposed scheme decreases
when $\Delta b$ increases, which means that our proposed scheme can create more coded-multicasting opportunities with
a relaxed delay requirement.
Furthermore,  the larger $\Delta b$ is, the more the decrease of the fronthaul load of our proposed scheme is  when compared with that of the uncoded caching scheme.
And the performance gap between the fronthaul load of our proposed scheme and that of the Maddah-Ali-Niesen's decentralized
synchronous coded caching scheme is smaller when $\Delta b$ is larger.
The reason for the above results is that a larger  $\Delta b$  leads to   a fewer number of  the partitioned subsets.
Moreover,
as $\Delta b$ determines the upper bound of the request delay,
it can be set to a relatively small value  in delay-sensitive scenarios and adjusted flexibly to achieve the  load-delay tradeoff in other scenarios.

\section{Conclusions}

In this paper, we have proposed a decentralized asynchronous coded caching scheme for the online case in F-RAN where users asynchronously request contents with the maximum request delay.
Our proposed  scheme provides asynchronous and synchronous transmission methods to fulfill the  delay requirements of different practical scenarios.
The simulation results have shown that more coded-multicasting opportunities can be created when the maximum request delay increases in asynchronous request scenarios.

\section*{Acknowledgments}

This work was supported in part by
the Research Fund of the State Key Laboratory of
Integrated Services Networks (Xidian University) under grant ISN19-10,
the Research Fund of the Key Laboratory of Wireless Sensor Network $\&$ Communication (Shanghai Institute of Microsystem and Information Technology, Chinese Academy of Sciences) under grant 2017002,
the Hong Kong, Macao and Taiwan Science $\&$ Technology Cooperation Program of China under grant 2014DFT10290,
the Ericsson and SEU Cooperation Project under grant 8504000335,
the National Basic Research Program of China
(973 Program) under grant 2012CB316004,
and the U.K. Engineering and Physical Sciences Research Council under Grant EP/K040685/2.

\bibliographystyle{IEEEtran}
\bibliography{codedcaching}

\end{document}